\documentclass[12pt]{article}
\usepackage{mitpress}
\usepackage{amsmath}
\usepackage{graphicx}
\usepackage{amsfonts}
\usepackage{amssymb}

\newdimen\dummy
\dummy=\oddsidemargin
\begin{document}

\title{Two-state teleportation}
\author{L. Henderson, L. Hardy and V. Vedral\\Centre for Quantum Computation, Clarendon Laboratory, University of Oxford,\\Parks Road OX1 3PU}
\date{\today}
\maketitle
\begin{abstract}
Quantum teleportation with additional a priori information about the input
state achieves higher fidelity than teleportation of a completely unknown
state. However, perfect teleportation of two non-orthogonal input states
requires the same amount of entanglement as perfect teleportation of an
unknown state, namely one ebit. We analyse how well two-state teleportation
can be achieved using every degree of pure-state entanglement. We find the
highest fidelity of `teleportation' that can be achieved with only classical
communication but no shared entanglement. A two-state telecloning scheme is constructed.
\end{abstract}

\section{Introduction}

Transmission of quantum states can be accomplished either by direct sending of
qubits, or by transmission of classical bits where the sender and receiver
share entanglement. In schemes for quantum teleportation it has been shown
that the transmission of two classical bits together with the use of one ebit
achieves the same results as sending one qubit, \cite{Bennett93}.

If the state to be teleported is completely unknown, the fact that the amount
of entanglement between two separated subsystems may not increase under local
operations means that faithful teleportation cannot be achieved without one
full unit of entanglement. The argument goes as follows. Alice's particle is
initially in an unknown state, which could be a mixed state due to
entanglement with another particle $R$ at Alice's end. After the
teleportation, the entanglement between Alice's particle and $R$ is
transferred to an entanglement between Bob's particle and $R$, by entanglement
swapping, \cite{Zukowski93}. The original entangled channel between Alice and
Bob is completely destroyed. Local operations and classical communication
cannot increase the entanglement between Alice and Bob. Therefore the original
entanglement in the channel must be at least as high as the final entanglement
between Bob's particle and $R$. However the initial state of Alice's particle
is completely unknown, it may be a maximally mixed state, arising because
Alice's particle is maximally entangled to another particle, $R$. This would
make the final entanglement between Bob and $R$ maximal. Therefore the initial
entanglement in the channel must be maximal also, \cite{Plenio98}.

On the other hand, if Alice knows exactly what state she has, there is no need
for any entanglement to reliably transmit the state. She simply sends Bob
classical information saying which state it is, and he prepares it himself.

Between the two extremes of Alice possessing no prior information of the
state, and Alice possessing full information, she may have some prior
knowledge. For example, she may receive her qubits from a known ensemble
$\varepsilon=\{\left|  \phi_{x}\right\rangle ,p_{x}\}$ of states $\left|
\phi_{x}\right\rangle $ with probability $p_{x}$.We consider the situation
where Alice knows that a preparer of quantum states provides her with one of
two non-orthogonal states, say $\left|  \psi_{1}\right\rangle $ $=\cos
(\frac{\theta}{2})\left|  0\right\rangle +\sin(\frac{\theta}{2})\left|
1\right\rangle $ and $\left|  \psi_{2}\right\rangle $ $=\sin(\frac{\theta}%
{2})\left|  0\right\rangle +\cos(\frac{\theta}{2})\left|  1\right\rangle $
with equal probabilities. Alice then knows almost everything about the state.
In effect she has to transmit one bit of information to Bob saying which of
the two states she has. Is it possible to teleport the quantum state in this
case, using less than the full unit of entanglement required when the state is
completely unknown?

It turns out, rather surprisingly, that it is not possible and that a full
unit of entanglement is needed even for teleportation of only two states. This
is shown in section \ref{sec:ent}. In section \ref{sec:class}, we find the
upper bound for the fidelity of sending the state with no entanglement. In
section \ref{sec:methods}, we consider teleportation using a non-maximally
entangled channel. We make some connections between teleportation and cloning
in section \ref{sec:cloning}, and adapt the telecloning scheme of Murao
\textit{et al.}, \cite{Murao99}, to the case of telecloning two non-orthogonal
states. The two-state telecloning state is now different from that for
universal telecloning. We find that the amount of entanglement required
between sender and recipients is now state-dependent.

\section{Schemes without \label{sec:class}entanglement}

For comparison, we first determine what fidelity of transmission can be
achieved without using any entanglement, only classical communication. Alice
measures her state and sends the result to Bob, who makes his best guess of
the state based on this information. The fidelity of sending the state
$\left|  \psi\right\rangle $ is defined as%
\begin{equation}
F_{cl}(\left|  \psi\right\rangle )=\sum_{i=1}^{n}P(i|\psi)|\langle\psi
|\alpha_{i}\rangle|^{2} \label{eq:fid}%
\end{equation}
where $P(i|\psi)=$ $\langle\psi|A_{i}\left|  \psi\right\rangle $ is the
probability of Alice obtaining the result corresponding to the positive
operator $A_{i}$ out of $n$ possible outcomes of the POVM $\{A_{i}\}$ where
$\sum_{i=1}^{n}A_{i}=1$. The state $|\alpha_{i}\rangle$ is Bob's guess, given
outcome $i$.

When the input state is completely unknown, the average of the fidelity over
an even distribution of all states on the Bloch sphere is taken. It has been
shown that the average fidelity over all states is \cite{Massar95}
\begin{align*}
F_{cl}  &  =\int\sum_{i=1}^{n}P(i|\psi)|\langle\psi|\alpha_{i}\rangle
|^{2})d\Omega\\
&  =\frac{2}{3}%
\end{align*}
In this case, Alice may make an orthogonal measurement in any direction, and
it is optimal for Bob to prepare the state corresponding to Alice's result.

On the other hand, when Alice's state is drawn from an ensemble of two states,
$\{\left|  \psi_{1}\right\rangle ,\left|  \psi_{2}\right\rangle \}$ with equal
probabilities, the fidelity
\[
F_{cl}(\{\left|  \psi_{1}\right\rangle ,\left|  \psi_{2}\right\rangle
\})=\frac{1}{2}\sum_{i=1}^{n}\sum_{j=1}^{2}P(i|\psi_{j})|\langle\psi
_{j}|\alpha_{i}\rangle|^{2}%
\]
is much higher. This is the case we consider in this paper.

We first calculate the fidelity in the case where Bob simply prepares a
guessed state corresponding to one of the two input states, $\left|
\alpha_{i}\right\rangle \in\{\left|  \psi_{1}\right\rangle ,\left|  \psi
_{2}\right\rangle $ $\}$, for all $i=1...n$. Then the fidelity is limited only
by the errors Alice makes in measuring, due to the fact that the signal states
are non-orthogonal. We employ previous results on distinguishing two states.
These results have been derived with respect to two different ways of
characterising distinguishability. The states may either be distinguished so
as to minimise the probability of error in guessing the right state, or by
using an `unambiguous' measurement, which has no probability of error, but
which sometimes yields no information about the state.

It has been shown, \cite{Helstrom76}, that the smallest attainable probability
of error in distinguishing two states is
\[
P_{e}=\frac{1}{2}-\frac{1}{4}Tr(\left|  \rho_{1}-\rho_{0}\right|  )
\]
For two pure states $\left|  \psi_{1}\right\rangle $ and $\left|  \psi
_{2}\right\rangle $, the minimal probability of error may be derived from the
unitary evolution of the unknown state and an ancilla qubit, initially in the
state $\left|  0\right\rangle _{A}$, on which a projective measurement will be
performed in the $\{\left|  0\right\rangle ,\left|  1\right\rangle \}$ basis:%

\begin{align}
\left|  0\right\rangle _{A}\left|  \psi_{1}\right\rangle  &  \rightarrow
\sqrt{1-P_{e}}\left|  0\right\rangle _{A}\left|  \psi_{1}\right\rangle
+\sqrt{P_{e}}\left|  1\right\rangle _{A}\left|  \psi_{2}\right\rangle
\label{eq:unit}\\
\left|  0\right\rangle _{A}\left|  \psi_{2}\right\rangle  &  \rightarrow
\sqrt{P_{e}}\left|  0\right\rangle _{A}\left|  \psi_{1}\right\rangle
+\sqrt{1-P_{e}}\left|  1\right\rangle _{A}\left|  \psi_{2}\right\rangle
\nonumber
\end{align}
If the ancilla is measured in the state $\left|  0\right\rangle _{A}$, we
conclude the state is $\left|  \psi_{1}\right\rangle $ and if \ $\left|
1\right\rangle _{A}$, then we conclude $\left|  \psi_{2}\right\rangle $. The
requirement that this evolution be unitary gives
\[
P_{e}=\frac{1}{2}(1\pm\sqrt{1-|\langle\psi_{1}|\psi_{2}\rangle|^{2}})
\]
For two pure states $\left|  \psi_{1}\right\rangle $ $=\cos\frac{\theta}%
{2}\left|  0\right\rangle +\sin\frac{\theta}{2}\left|  1\right\rangle $ and
$\left|  \psi_{2}\right\rangle $ $=\sin\frac{\theta}{2}\left|  0\right\rangle
+\cos\frac{\theta}{2}\left|  1\right\rangle ,$ this is given by
\begin{equation}
P_{e}=\frac{1}{2}(1-\cos\theta)\nonumber
\end{equation}
If $\theta=0$, the two states are orthogonal and the probability of error is
zero. If no error is made, Bob prepares Alice's state with perfect fidelity.
If Alice makes an error, there is still some overlap with the correct state,
given by $\sin^{2}\theta$. The fidelity is therefore
\begin{align}
F  &  =(1-P_{e}).1+P_{e}\sin^{2}\theta\label{eq:minprob}\\
&  =1-\frac{1}{2}(1-\cos\theta)\cos^{2}\theta\nonumber
\end{align}
For orthogonal states, $\theta=0$, $F_{1}=1$. For maximally non-orthogonal
states with $\theta=\frac{\pi}{4}$, $F_{1}=0.927.$

An alternative strategy is to construct a POVM which distinguishes the two
outcomes, $\left|  \psi_{1}\right\rangle $ and $\left|  \psi_{2}\right\rangle
$, with no probability of error, but has a third outcome where the state is
completely unknown. Then the maximum probability of a successful outcome is
\cite{Ivanovic87}, \cite{Dieks88}, \cite{Peres88} \label{eq:probsuccess}which
is $P_{s}=1-\sin\theta$ in our case.
\begin{figure}
[ptb]
\begin{center}
\includegraphics[
height=3.5622in,
width=4.83in
]%
{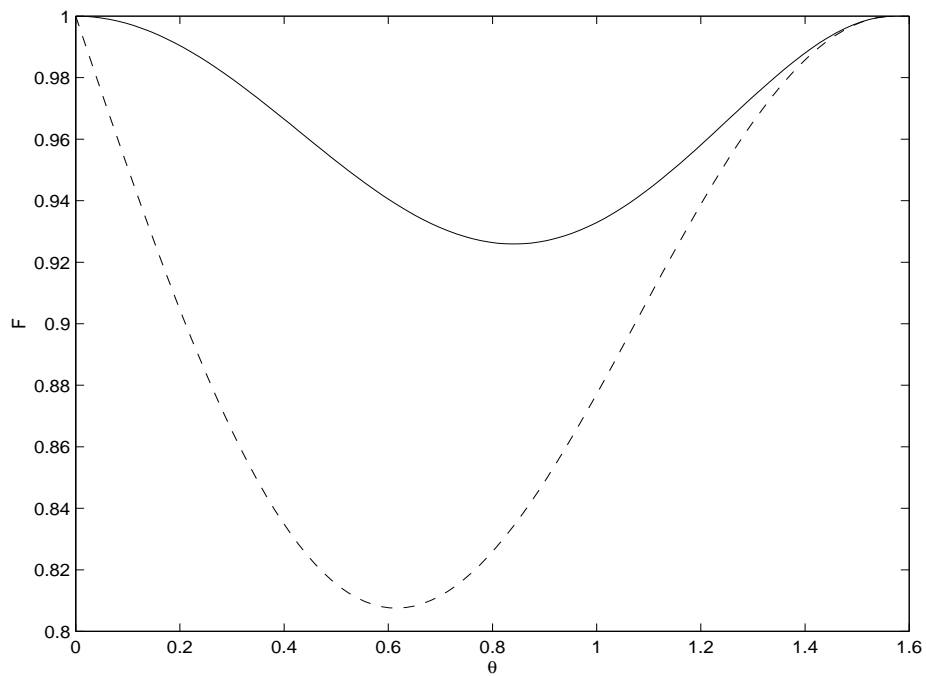}%
\caption{Fidelity when Bob guesses one of the two input states. The solid line
corresponds to the measurement which minimises Alice's probability of error,
Eq. (\ref{eq:minprob}), and the dashed line to a measurement giving
unambiguous discrimination of the two states, Eq. (\ref{eq:unamb}). }%
\label{fig1}%
\end{center}
\end{figure}
If the `don't know' outcome is obtained, Bob chooses at random which state to
prepare. In half the cases, he succeeds. If he fails, there is still an
overlap with the correct state. Therefore the fidelity is
\begin{equation}
F=1-\frac{1}{2}\sin\theta+\frac{1}{2}\sin^{3}\theta\label{eq:unamb}%
\end{equation}
This fidelity is always lower than that achieved by minimising the probability
of error, (see Fig. (\ref{fig1})).

\begin{figure}
[ptb]
\begin{center}
\includegraphics[
trim=0.000000in 0.000000in -0.339812in 0.099327in,
height=3.563in,
width=4.8282in
]%
{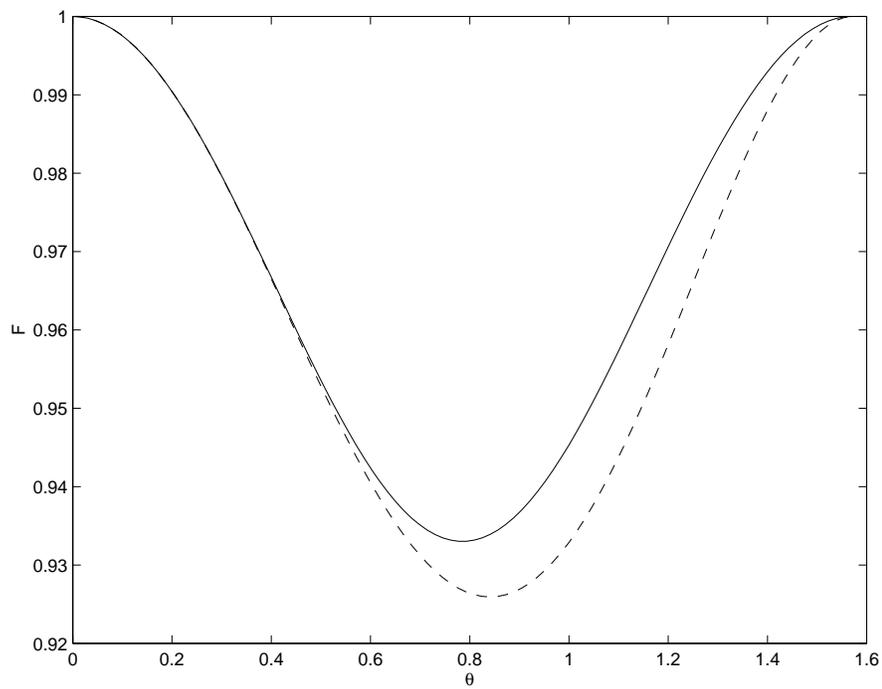}%
\caption{Fidelity when Alice minimises her probability of error. The dashed
line shows the case where Bob prepares the state she specifies, Eq.
(\ref{eq:minprob}), and the solid line the case where he optimises his guess,
Eq. (\ref{eq:fidtwo}).}%
\label{fig2}%
\end{center}
\end{figure}

However the strategy where Alice minimises her probability of error and Bob
prepares one of the input states is not optimal. It is possible to achieve a
higher fidelity if Bob prepares a guess which has a slightly higher overlap
with the other state to take into account the possibility that Alice makes an
error. Alice still makes the measurement which minimises her probability of
error. For the two states $\left|  \psi_{1}\right\rangle $ $=\cos(\frac
{\theta}{2})\left|  0\right\rangle +\sin(\frac{\theta}{2})\left|
1\right\rangle $ and $\left|  \psi_{2}\right\rangle $ $=\sin(\frac{\theta}%
{2})\left|  0\right\rangle +\cos(\frac{\theta}{2})\left|  1\right\rangle $
this is a projection onto $\left|  0\right\rangle $ or $\left|  1\right\rangle
$. The positive operators of the POVM to be performed are $A_{1}=\left|
0\right\rangle \langle0|$ and $A_{2}=\left|  1\right\rangle \langle1|$ and the
corresponding probabilities are
\begin{align*}
p(1|\psi_{1})  &  =|\left\langle 0|\psi_{1}\right\rangle |^{2}=\cos^{2}%
(\frac{\theta}{2})\\
p(2|\psi_{1})  &  =|\left\langle 1|\psi_{1}\right\rangle |^{2}=\sin^{2}%
(\frac{\theta}{2})\\
p(1|\psi_{2})  &  =|\left\langle 0|\psi_{2}\right\rangle |^{2}=\sin^{2}%
(\frac{\theta}{2})\\
p(2|\psi_{2})  &  =|\left\langle 0|\psi_{2}\right\rangle |^{2}=\cos^{2}%
(\frac{\theta}{2})
\end{align*}
The fidelity is
\begin{align*}
F_{cl}  &  =\frac{1}{2}(p(1|\psi_{1})|\left\langle \alpha|\psi_{1}%
\right\rangle |^{2}+p(2|\psi_{1})|\left\langle \beta|\psi_{1}\right\rangle
|^{2}\\
&  +p(1|\psi_{2})|\left\langle \alpha|\psi_{2}\right\rangle |^{2}+p(2|\psi
_{2})|\left\langle \beta|\psi_{2}\right\rangle |^{2})
\end{align*}
where $\left|  \alpha\right\rangle $ and $\left|  \beta\right\rangle $ are
Bob's guessed states. Assuming that the fidelity must be the same under
interchange of the two states, and that the guessed states share the same
symmetry as the input states, so that $|\left\langle \alpha|\psi
_{1}\right\rangle |^{2}=|\left\langle \beta|\psi_{2}\right\rangle |^{2}$, and
$|\left\langle \beta|\psi_{1}\right\rangle |^{2}=|\left\langle \alpha|\psi
_{2}\right\rangle |^{2}$, the fidelity becomes
\begin{align}
F_{cl}  &  =p(1|\psi_{1})|\left\langle \alpha|\psi_{1}\right\rangle
|^{2}+p(2|\psi_{1})|\left\langle \beta|\psi_{1}\right\rangle |^{2}%
\label{eq:fidtwo}\\
&  =\cos^{2}\frac{\theta}{2}\cos^{2}(\frac{\theta-\alpha}{2})+\sin^{2}%
\frac{\theta}{2}\sin^{2}(\frac{\theta+\alpha}{2})\nonumber
\end{align}
Differentiating with respect to the choice of guessed angle $\alpha$ gives%
\[
\frac{\partial F_{cl}}{\partial\alpha}=p(1|\psi_{1})\sin(\theta-\alpha
)+p(2|\psi_{1})\sin(\theta+\alpha)
\]
We find the maximum value of $F_{cl}$ by setting $\frac{\partial F_{cl}%
}{\partial\alpha}=0$. The angle which gives a maximum is
\[
\alpha=\tan^{-1}\left(  \frac{\sin\theta}{\cos^{2}\theta}\right)
\]
Substituting into Eq. (\ref{eq:fidtwo}) gives the fidelity plotted in Fig.
(\ref{fig2}). Notice that this fidelity, unlike the fidelity of the other
strategies, is symmetrical about $\theta=\frac{\pi}{4}$.
This result coincides with the following expression derived by Fuchs and
Peres, \cite{Fuchs95}, in the context of eavesdropping,%
\[
F_{cl}=\frac{1}{2}(1+\sqrt{1-|\langle\psi_{1}|\psi_{2}\rangle|^{2}%
+|\langle\psi_{1}|\psi_{2}\rangle|^{4}})
\]
In this scenario, Alice tries to communicate to Bob one of a set of
non-orthogonal states, which is intercepted by Eve. Eve wants to extract as
much information as possible from a measurement on the state, and at the same
time to prepare a new state with as high fidelity as possible with Alice's
original state so as to deceive Bob. Eve performs the dual function of Alice
as measurer and Bob as preparer in our scheme where Alice and Bob are
connected only by a classical channel. It is plausible that for Bob to
maximise the fidelity, he should have maximum information about the state and
that Alice should also maximise her information by making the measurement
which minimises the probability of error. This suggests that the fidelity of
Eq. (\ref{eq:fidtwo})\ is optimal. The symmetry about $\theta=\frac{\pi}{4}$
may indicate optimality since all the less efficient strategies that we
investigated do not possess this symmetry. Fuchs and Peres give further
numerical and plausibility arguments in support of the optimality of this fidelity.

Up till now the discussion has focussed on the situation where Alice and Bob
communicate only by a classical channel. We now consider how shared
entanglement can improve the fidelity of teleportation.

\section{Use of entanglement\label{sec:ent}}

If Alice and Bob share only one entangled pair, perfect two-state
teleportation cannot be achieved without a full unit of entanglement. By
contrast, in the asymptotic case with many copies of the state and many
entangled pairs, perfect teleportation may be achieved with less than one full
unit of entanglement for each qubit communicated.

\subsection{Single channel case\label{sect:single}}

We prove that it is not possible to teleport perfectly with less than one full
unit of entanglement, even if the state to be teleported comes from a known
ensemble of only two non-orthogonal states. Let the state to be teleported be
$\left|  \phi\right\rangle _{1}$, and the entangled channel $\left|
\psi\right\rangle _{23}$. Then the initial state of the three particles may be
written as%
\[
\left|  \phi\right\rangle _{1}\left|  \psi\right\rangle _{23}=\sum_{k}%
c_{k}^{\phi}\left|  k\right\rangle _{12}U_{k}^{-1}\left|  \phi\right\rangle
_{3}%
\]
where the coefficient $c_{k}^{\phi}$ may depend on the initial state $\phi$.
The state has been expanded as a bipartite decomposition of the first two
particles versus the third, where the orthonormal basis of the first two
particles is given by $\{\left|  k\right\rangle _{12}\}$ and the corresponding
states of the third particle are $U_{k}^{-1}\left|  \phi\right\rangle _{3}$,
not necessarily orthogonal. Any general teleportation scheme must be of this
form. The state can be transformed unitarily as%
\begin{equation}
U(\left|  \phi\right\rangle _{1}\left|  \psi\right\rangle _{23})=(\sum
_{k}c_{k}^{\phi}\left|  k\right\rangle _{12})\left|  \phi\right\rangle _{3}%
\end{equation}
by the controlled unitary operation $U_{k}$ on the third particle. Let
$|A(\phi)\rangle_{12}=(\sum_{k}c_{k}^{\phi}\left|  k\right\rangle _{12})$ and
consider two input states, $\left|  \phi\right\rangle $ and $\left|
\phi^{\prime}\right\rangle $. By taking the overlap of Eq. (\ref{eq:unit})
with a similar equation for $\left|  \phi^{\prime}\right\rangle $, we obtain%
\[
_{1}\langle\phi^{\prime}\left|  \phi\right\rangle _{1}=(_{12}\langle
A(\phi^{\prime})|A(\phi)\rangle_{12})(_{3}\langle\phi^{\prime}\left|
\phi\right\rangle _{3})
\]
Since%
\[
_{1}\langle\phi^{\prime}\left|  \phi\right\rangle _{1}=_{3}\langle\phi
^{\prime}\left|  \phi\right\rangle _{3}%
\]
it follows that either $\langle\phi^{\prime}\left|  \phi\right\rangle =0$, or
$_{12}\langle A(\phi^{\prime})|A(\phi)\rangle_{12}=1$. If $\ \langle
\phi^{\prime}\left|  \phi\right\rangle =0$, the two input states are
orthogonal and perfect teleportation can be achieved without the use of any
entanglement at all, since an exact measurement to distinguish the states can
be performed. The vectors $\left|  A(\phi)\right\rangle _{12}$ and $\left|
A(\phi^{\prime})\right\rangle _{12}$ are normalised. Hence, if $_{12}\langle
A(\phi^{\prime})|A(\phi)\rangle_{12}=1$, then $\left|  A(\phi)\right\rangle
_{12}=$ $\left|  A(\phi^{\prime})\right\rangle _{12}$ and consequently the
coefficients $c_{k}^{\phi}$ must be independent of the input state $\phi$, so
that $c_{k}^{\phi}=c_{k}^{\phi^{\prime}}$. Therefore the probability of
obtaining the result $k$ is independent of the input state.

Any state to be teleported can be written as a linear combination of the
states $\left|  \phi\right\rangle $ and $\left|  \phi^{\prime}\right\rangle $%
\[
\left|  \psi\right\rangle =a\left|  \phi\right\rangle +b\left|  \phi^{\prime
}\right\rangle
\]
If both $\left|  \phi\right\rangle $ and $\left|  \phi^{\prime}\right\rangle $
can be teleported perfectly by the same operation, there exists a unitary
transformation $U$ such that%
\[
U(\left|  \phi\right\rangle _{1}\left|  \psi\right\rangle _{23})=(\sum
_{k}c_{k}\left|  k\right\rangle _{12})\left|  \phi\right\rangle _{3}%
\]
and%
\[
U(\left|  \phi^{\prime}\right\rangle _{1}\left|  \psi\right\rangle
_{23})=(\sum_{k}c_{k}\left|  k\right\rangle _{12})\left|  \phi^{\prime
}\right\rangle _{3}%
\]
where we have shown that the coefficients $c_{k}$ do not depend on the input
state. Therefore
\[
U(a\left|  \phi\right\rangle _{1}+b\left|  \phi^{\prime}\right\rangle
_{1})\left|  \psi\right\rangle _{23}=\sum_{k}c_{k}\left|  k\right\rangle
_{12}(a\left|  \phi\right\rangle _{3}+b\left|  \phi^{\prime}\right\rangle
_{3})
\]
and so any state can be teleported perfectly. This would mean it were possible
to perfectly teleport a maximally mixed state. By the arguments of the
introduction this would require a full unit of entanglement.

\subsection{Asymptotic case\label{sec:asymp}}

Alice's qubit is an equally weighted mixture of the two possible input states
and so can be described by the density matrix
\[
\rho=\frac{1}{2}(\left|  \psi_{1}\right\rangle \langle\psi_{1}|+\left|
\psi_{2}\right\rangle \langle\psi_{2}|)
\]
Now if Alice possesses a large number $n$ of copies of the qubit, she may use
Schumacher compression, \cite{Schumacher95}, to compress the same information
into $nS(\rho)$ qubits, where $S(\rho)=-tr(\rho\log\rho)$ is the Von Neumann
entropy of the qubit $\rho$. If $\theta=0$, and the two states are orthogonal,
$S(\rho)=\log2=1$. \ This is the only case where no compression is possible.
For two maximally non-orthogonal states, with $\theta=\frac{\pi}{4}$,
$S(\rho)\approx0.907$ and transmission requires $0.907$ ebits per qubit of information.

If Alice and Bob share a large number $m$ of non-maximally entangled pairs in
the state $\rho_{AB}$, with $\rho_{A}=Tr_{B}(\rho_{AB})$, they may distill
$mS(\rho_{A})$ maximally entangled pairs using only local operations and
classical communication, \cite{Bennett96},\cite{Bennett96prl}. The quantity
$S(\rho_{A})$ denotes the amount of entanglement in the shared pairs and for a
maximally entangled state, $S(\rho_{A})=1$. The amount of entanglement
$S(\rho_{A})$ required per qubit of information sent by Alice is $S(\rho
_{A})=\frac{n}{m}S(\rho)$, which may be less than one in the limit of large
$m$ and $n$, when the input states are non-orthogonal. Clearly then, the
asymptotic case is different from the situation where only single copies of
the states are available.

\section{Teleportation through a non-maximally entangled
channel\label{sec:methods}}

Given that when Alice and Bob share only one non-maximally entangled channel
it is not possible to perform two-state teleportation perfectly, we would like
to know how high a fidelity can be achieved. Below, we compare several
different strategies, however it is still an open question what the most
optimal scheme would be.

If we apply the standard teleportation procedure, sending the initial state
\[
\left|  \psi\right\rangle _{1}=\cos\frac{\theta}{2}\left|  0\right\rangle
+\sin\frac{\theta}{2}\exp(i\phi)\left|  1\right\rangle
\]
through the non-maximally entangled channel%
\[
\left|  \psi\right\rangle _{23}=\alpha\left|  00\right\rangle +\beta\left|
11\right\rangle
\]
then the initial state of the three particles may be written as%

\begin{align}
\left|  \psi\right\rangle _{123}  &  =\frac{1}{\sqrt{2}}(\left|  \phi
^{+}\right\rangle (\alpha\cos\frac{\theta}{2}\left|  0\right\rangle +\beta
\sin\frac{\theta}{2}e^{i\phi}\left|  1\right\rangle )\label{eq:bellbasis}\\
&  +\left|  \phi^{-}\right\rangle (\alpha\cos\frac{\theta}{2}\left|
0\right\rangle -\beta\sin\frac{\theta}{2}e^{i\phi}\left|  1\right\rangle
)\nonumber\\
&  +\left|  \psi^{+}\right\rangle (\alpha\sin\frac{\theta}{2}e^{i\phi}\left|
0\right\rangle +\beta\cos\frac{\theta}{2}\left|  1\right\rangle )\nonumber\\
&  +\left|  \psi^{-}\right\rangle (-\alpha\sin\frac{\theta}{2}e^{i\phi}\left|
0\right\rangle +\beta\cos\frac{\theta}{2}\left|  1\right\rangle ))\nonumber
\end{align}
Without loss of generality, we assume that $\alpha$ and $\beta$ are real and
that $\alpha\leq\beta$. The fidelity is given by%
\[
F(\left|  \psi\right\rangle )=\sum_{i=1}^{4}p(i|\psi)|\left\langle \psi
|\alpha_{i}\right\rangle |^{2}%
\]
where $i$ is the index of the projections $A_{i}=\left|  \phi_{i}\right\rangle
\langle\phi_{i}|$ onto the four Bell states
\begin{align*}
\left|  \phi_{1}\right\rangle  &  =\left|  \phi^{+}\right\rangle \\
\left|  \phi_{2}\right\rangle  &  =\left|  \phi^{-}\right\rangle \\
\left|  \phi_{3}\right\rangle  &  =\left|  \psi^{+}\right\rangle \\
\left|  \phi_{4}\right\rangle  &  =\left|  \psi^{-}\right\rangle
\end{align*}
and $\left|  \alpha_{i}\right\rangle $ is Bob's normalised and corrected
outcome $\left|  \alpha_{i}\right\rangle $ given the measurement result $i$.
The probability of Alice measuring $\left|  \phi^{+}\right\rangle $ or
$\left|  \phi^{-}\right\rangle $, given the input state $\left|
\psi\right\rangle =\cos\frac{\theta}{2}\left|  0\right\rangle +\sin
\frac{\theta}{2}\exp(i\phi)\left|  1\right\rangle $ is%
\[
p(1|\psi)=p(2|\psi)=\frac{1}{2}(\alpha^{2}\cos^{2}\frac{\theta}{2}+\beta
^{2}\sin^{2}\frac{\theta}{2})
\]
and of measuring $\left|  \psi^{+}\right\rangle $ or $\left|  \psi
^{-}\right\rangle $ is%
\[
p(3|\psi)=p(4|\psi)=\frac{1}{2}(\alpha^{2}\sin^{2}\frac{\theta}{2}+\beta
^{2}\cos^{2}\frac{\theta}{2})
\]
The fidelity is then%
\[
F(\left|  \psi\right\rangle )=\cos^{4}\frac{\theta}{2}+\sin^{4}\frac{\theta
}{2}+\alpha\beta\sin^{2}\theta
\]
Averaged over all initial states, this gives%
\begin{align}
F_{av}  &  =\frac{1}{4\pi}\int_{0}^{2\pi}\int_{0}^{\pi}(\cos^{4}\frac{\theta
}{2}+\sin^{4}\frac{\theta}{2}+\alpha\beta\sin^{2}\theta)\sin\theta d\theta
d\phi\label{eq:diruni}\\
&  =\frac{2}{3}(1+\alpha\beta)\nonumber
\end{align}
It can be shown, using a result of the Horodeckis, \cite{Horodecki98}, that
the average fidelity given in Eq. (\ref{eq:diruni}) is optimal for any
teleportation scheme, whatever Alice's measurement or Bob's corrections. The
Horodeckis derive a general relation between the optimal fidelity of
teleportation $\ F_{tele}$ and the maximal singlet fraction $f$, defined
below, of the state used for teleportation%
\[
F_{tele}=\frac{2f+1}{3}%
\]
For the non-maximally entangled state $\alpha\left|  00\right\rangle
+\beta\left|  11\right\rangle $, the maximal singlet fraction is
\begin{align*}
f  &  =|\frac{1}{\sqrt{2}}(\langle00|+\langle11|)(\alpha\left|
00\right\rangle +\beta\left|  11\right\rangle )|^{2}\\
&  =\frac{1}{2}(1+2\alpha\beta)
\end{align*}
and hence the optimal fidelity of teleportation is given by Eq.
(\ref{eq:diruni}).

In the two-state case, where Alice has either $\left|  \psi_{1}\right\rangle $
$=\cos(\frac{\theta}{2})\left|  0\right\rangle +\sin(\frac{\theta}{2})\left|
1\right\rangle $ or $\left|  \psi_{2}\right\rangle $ $=\sin(\frac{\theta}%
{2})\left|  0\right\rangle +\cos(\frac{\theta}{2})\left|  1\right\rangle $
with equal probabilities, the fidelity is%
\begin{equation}
F=\cos^{4}\frac{\theta}{2}+\sin^{4}\frac{\theta}{2}+\alpha\beta\sin^{2}%
\theta\label{eq:dirtwo}%
\end{equation}
When $\left|  \psi_{1}\right\rangle $ and $\left|  \psi_{2}\right\rangle $ are
not orthogonal, the fidelity can only be unity if the channel is maximally
entangled, $\alpha=\beta=\frac{1}{\sqrt{2}}$.

Another strategy for teleportation is based on first purifying the channel.
Purification has some probability to convert the state to a maximally
entangled state, which can achieve perfect teleportation, and some probability
to fail so that no entanglement is shared, and Alice and Bob must revert to
the classical methods for sending the state with no shared entanglement. For a
single copy, the best purification is the `Procrustean' method,
\cite{Bennett96}, which has optimal efficiency $2\alpha^{2}$, \cite{Lo97}.
When the purification fails, Alice and Bob are left with a product state. The
input state is unaffected by purification, so Alice may employ the best
strategy for transmitting the state without entanglement. For a completely
unknown input state, the fidelity is $F_{cl}=\frac{2}{3}$, hence the fidelity is%

\begin{equation}
F=\frac{2}{3}(1+\alpha^{2}) \label{eq:purif}%
\end{equation}
Higher fidelities are achieved in the two-state case. Then the best fidelity
which may be achieved is
\begin{equation}
F=2\alpha^{2}+(1-2\alpha^{2})F_{cl} \label{eq:puriftwo}%
\end{equation}
where
\[
F_{cl}=\cos^{2}\frac{\theta}{2}\cos^{2}(\frac{\theta-\alpha}{2})+\sin^{2}%
\frac{\theta}{2}\sin^{2}(\frac{\theta+\alpha}{2})
\]
is the best measurement strategy with no entanglement with
\[
\alpha=\tan^{-1}\left(  \frac{\sin\theta}{\cos^{2}\theta}\right)
\]

\begin{figure}
[ptb]
\begin{center}
\includegraphics[
trim=0.000000in 0.000000in -0.466126in 0.000000in,
height=3.563in,
width=4.8282in
]%
{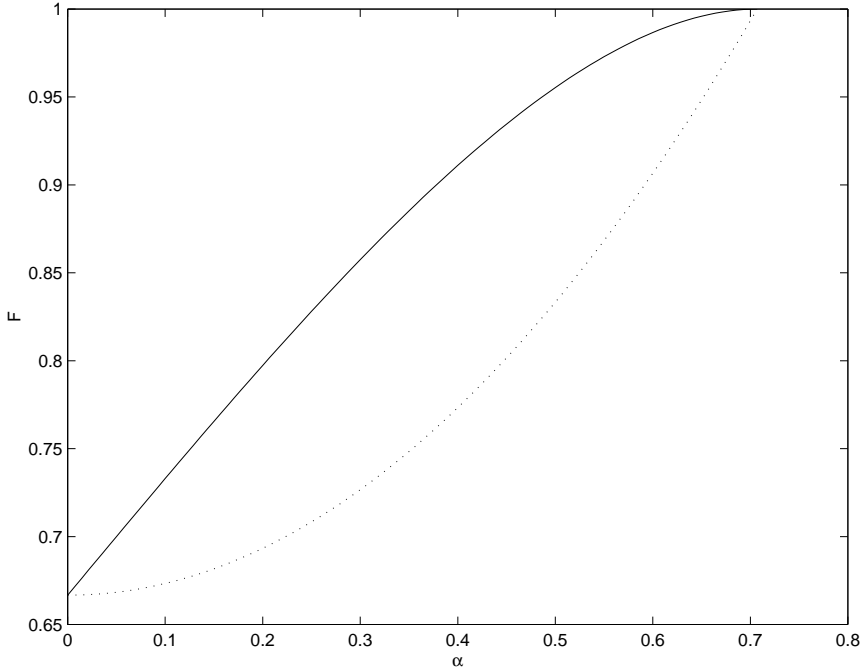}%
\caption{Teleportation through a non-maximally entangled channel for unknown
state. The dotted line shows the purification method, Eq. (\ref{eq:purif}),
the solid line the direct method, Eq. (\ref{eq:diruni}).}%
\label{fig:uni}%
\end{center}
\end{figure}

\begin{figure}
[ptb]
\begin{center}
\includegraphics[
trim=0.000000in 0.000000in -0.466126in 0.000000in,
height=3.563in,
width=4.8282in
]%
{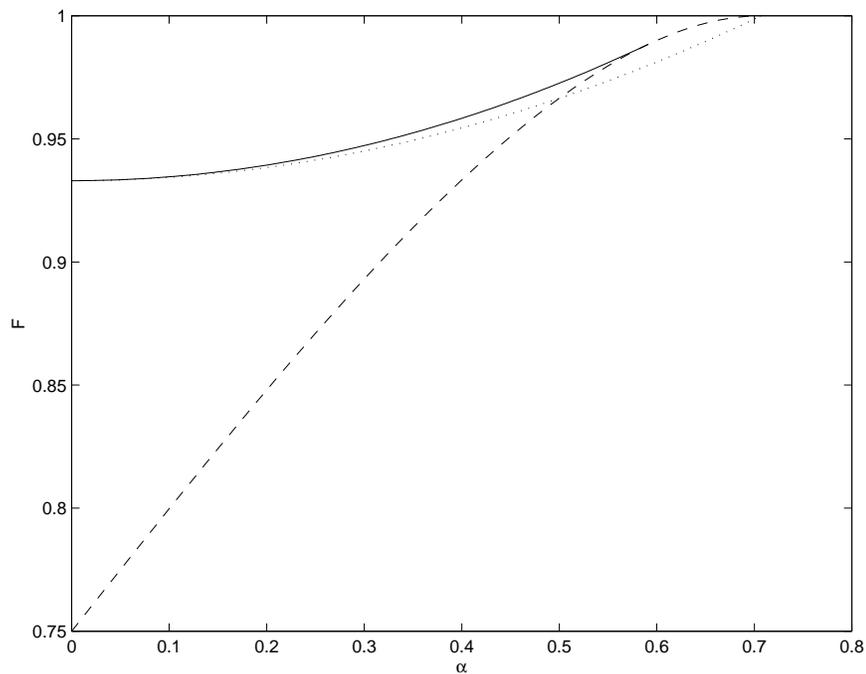}%
\caption{Teleportation through a non-maximally entangled channel for two
states with $\theta=\frac{\pi}{4}$. The dotted line shows the purification
method, Eq. (\ref{eq:puriftwo}), the dashed line the direct method using the
standard corrections, Eq. (\ref{eq:dirtwo}), and the solid line the optimal
combination of the two methods, Eq. (\ref{eq:comb}).}%
\label{fig:comb}%
\end{center}
\end{figure}

For a completely unknown input state, teleporting directly through the
non-maximally entangled channel is always better than the strategy based on
purification, Eq. (\ref{eq:purif}), since $\alpha\leq\beta$, see Fig.
(\ref{fig:uni}).
For two input states, on the other hand, the fidelities of the different
methods are plotted in Fig. (\ref{fig:comb}). The direct method is no longer
always better than the purification method, though it is better when the
entanglement in the channel is high, in which case it approaches the average
fidelity. For low entanglement, the efficiency of the direct method falls off
steeply and becomes worse even than the classical strategy without entanglement.

For a completely unknown state, teleportation via either strategy is always
better than the classical method of measuring and communicating the result.
However, when there are only two possible input states, a large amount of
information may be gained just by Alice measuring the state she has. It turns
out that the fidelity which may be achieved by Alice measuring her state and
telling Bob the result classically is higher than a direct teleportation, if
the channel has low entanglement. On the other hand, when the channel is first
purified, it is possible to take advantage of the high classical fidelity, by
employing the classical strategy when the purification fails. It is possible
to do this because it is known when the purification has failed. Hence in the
two-state case, the purification method is better for low entanglement than
the direct method.

In the two-state case, it is not known what is the optimal teleportation
scheme is. The best bound we have found is based on a combination of the
direct and purification methods. This may be achieved by Alice partially
purifying the entangled channel, $\alpha\left|  00\right\rangle +\beta\left|
11\right\rangle $ to a more entangled channel, $\alpha^{\prime}\left|
00\right\rangle +\beta^{\prime}\left|  11\right\rangle $, where $\alpha
^{\prime}\geq\alpha$. The probability of succeeding in this purification is
$P_{s}=\left(  \frac{\alpha}{\alpha^{\prime}}\right)  ^{2}$. If the
purification succeeds, the direct method may be employed on the more entangled
state. If it fails, the best classical strategy must be employed. Hence the
fidelity is given by%
\begin{equation}
F=\left(  \frac{\alpha}{\alpha^{\prime}}\right)  ^{2}F_{dir}(\alpha^{\prime
})+(1-\left(  \frac{\alpha}{\alpha^{\prime}}\right)  ^{2})F_{class}
\label{eq:comb}%
\end{equation}
For a particular non-maximally entangled channel, $\alpha$, this fidelity is
maximised by purifying to a particular channel characterised by $\alpha
^{\prime}$.

\section{Relation to telecloning\label{sec:cloning}}

Limitations on the fidelity of teleportation can be related to limitations on
the fidelity of cloning non-orthogonal quantum states. When a perfect
teleportation is achieved, there should be no information about the state left
on Alice's side which would enable her to construct any approximate copy of
the state in addition to the perfectly teleported state. Teleportation using a
maximally entangled pair achieves perfect fidelity, and the measurement on
Alice's side provides no information since the probability of obtaining the
different measurement outcomes is independent of the input state. This was
also indicated by Nielsen and Caves, \cite{Nielsen97}, who showed that
teleportation is a special case of reversing a quantum measurement, and that a
necessary condition for reversibility of a general quantum operation is that
no information about the prior state be obtainable from the measurement. On
the other hand, if the channel is not maximally entangled, perfect
teleportation cannot be achieved and Alice's measurement may provide some
information about the input state. We have seen that when there is no
entanglement in the channel at all, the optimal strategy is for Alice to
extract as much information as possible from her measurement. The measurement
result may then be used to prepare an arbitrary number, $M$, of identical
imperfect copies of the original state with fidelity given by Eq.
(\ref{eq:fid}). This type of cloning has been called `classical cloning',
\cite{Gisin97}, to distinguish it from the more general operation of quantum
cloning which is based on unitary evolution of the input with an ancilla.
Quantum cloning can achieve higher fidelities than classical cloning for a
finite number of copies $M$. The process of quantum cloning allows the use of
more entanglement than classical cloning since the ancilla may remain
entangled to the input and the copies, which may also be entangled to one
another. For two-state teleportation through a non-maximally entangled
channel, there is a trade-off between the classical cloning based on directly
measuring the input state, and the fidelity that can be achieved by
teleportation based on the entanglement. The exact relation between the
constraints on sharing information amongst copies in cloning and in
teleportation is a topic for further research. However, one way in which the
relationship between cloning and teleportation may be pursued is through a
combination of the two procedures in `telecloning'. We now investigate the
effect of a priori information on this protocol.

\subsection{State-dependent telecloning\label{sec:teleclone}}

Teleportation has been combined with optimal universal cloning from one to $M$
\ copies, \cite{Murao99}. This is achieved by performing the usual
teleportation protocol but with the entangled channel being a multiparticle
entangled state, called the `telecloning' state. For $M=2$, the telecloning
state is a $4$-qubit state%
\[
\left|  \psi_{TC}\right\rangle =\frac{1}{\sqrt{2}}(\left|  0\right\rangle
\left|  \phi_{0}\right\rangle +\left|  1\right\rangle \left|  \phi
_{1}\right\rangle )\label{eq:tcstate}%
\]
where $\left|  \phi_{0}\right\rangle $ and $\left|  \phi_{1}\right\rangle $
are the optimal cloning states produced by acting with the optimal cloning
transformation $U_{12}$ on $\left|  0\right\rangle $ and $\left|
1\right\rangle $ respectively,%
\begin{align*}
\left|  \phi_{0}\right\rangle  &  =U_{12}(\left|  0\right\rangle _{A}\left|
00\right\rangle )=\sqrt{\frac{2}{3}}\left|  0\right\rangle _{A}\left|
00\right\rangle +\sqrt{\frac{1}{6}}\left|  1\right\rangle _{A}(\left|
01\right\rangle +\left|  10\right\rangle )\\
\left|  \phi_{1}\right\rangle  &  =U_{12}(\left|  0\right\rangle _{A}\left|
10\right\rangle )=\sqrt{\frac{2}{3}}\left|  1\right\rangle _{A}\left|
11\right\rangle +\sqrt{\frac{1}{6}}\left|  0\right\rangle _{A}(\left|
01\right\rangle +\left|  10\right\rangle )
\end{align*}
where subscript $A$ denotes the ancilla. In the telecloning state, the first
two qubits and the `port' are held by Alice and the last two qubits belong to
two distant users, Bob and Claire. When the other qubits are traced over after
telecloning, these yield the optimal clones of Alice's input state. The total
amount of entanglement between Alice and the other users, given by the Von
Neumann entropy of the reduced density matrix after tracing over one side, was
found to be $\log(3)$, clearly less than the two units of entanglement
required if cloning is performed first and then the standard teleportation.

\begin{figure}
[ptb]
\begin{center}
\includegraphics[
height=3.0787in,
width=3.9072in
]%
{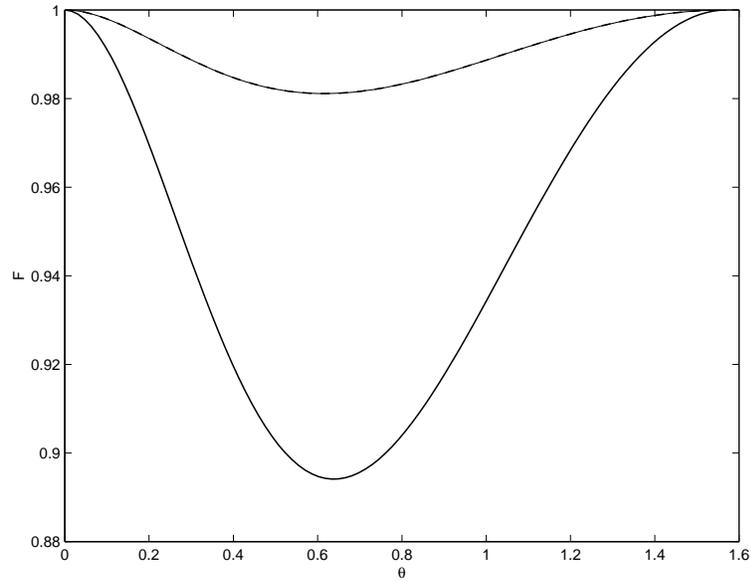}%
\caption{The global fidelity of the clones produced in the telecloning scheme,
(solid line), as compared to the optimal global fidelity for two-state
cloning, (dotted line).}%
\label{fig:newfil}%
\end{center}
\end{figure}

Adapting the telecloning scheme to communicating two states produces a
surprising result in terms of the amount of entanglement required. Bruss
\textit{et al.}, \cite{Bruss98}, have found the optimal cloning transformation
$U$ with respect to the global fidelity for two-state cloning from one copy to
two. An ancilla is not necessary. Following the same procedure as in the
universal case for constructing the telecloning state, we may add an ancilla,
giving the cloned states to be%
\begin{align*}
\left|  \phi_{0}\right\rangle  &  =U_{12}(\left|  0\right\rangle _{A}\left|
00\right\rangle )=a\left|  0\right\rangle _{A}\left|  00\right\rangle
+b\left|  1\right\rangle _{A}(\left|  01\right\rangle +\left|  10\right\rangle
)+c\left|  0\right\rangle _{A}\left|  11\right\rangle )\\
\left|  \phi_{1}\right\rangle  &  =U_{12}(\left|  0\right\rangle _{A}\left|
10\right\rangle )=c\left|  1\right\rangle _{A}\left|  00\right\rangle
+b\left|  0\right\rangle _{A}(\left|  01\right\rangle +\left|  10\right\rangle
)+a\left|  1\right\rangle _{A}\left|  11\right\rangle )
\end{align*}
where $a$, $b$ and $c$ depend on the overlap of the two states, as given in
the paper \cite{Bruss98}. The telecloning state is constructed just as before,
Eq. (\ref{eq:tcstate}). The ancilla is required in order that the recipients
may use the standard Pauli rotations to correct their state after they receive
the classical message from Alice. Notice however that $\left|  \phi
_{0}\right\rangle $ and $\left|  \phi_{1}\right\rangle $ are no longer the
optimal clones. The fidelity of cloning is shown in Figure (\ref{fig:newfil}).%

The reduced density matrix found by tracing the density matrix for the
telecloning state over Alice's two qubits is%
\[
\rho_{34}=\frac{1}{2}\left(
\begin{array}
[c]{cccc}%
a^{2}+b^{2}+c^{2} & 0 & 0 & 2a(b+c)\\
0 & b^{2} & 0 & 0\\
0 & 0 & b^{2} & 0\\
2a(b+c) & 0 & 0 & a^{2}+b^{2}+c^{2}%
\end{array}
\right)
\]%
\begin{figure}
[ptb]
\begin{center}
\includegraphics[
trim=0.000000in 0.000000in -0.466126in 0.000000in,
height=3.563in,
width=4.8282in
]%
{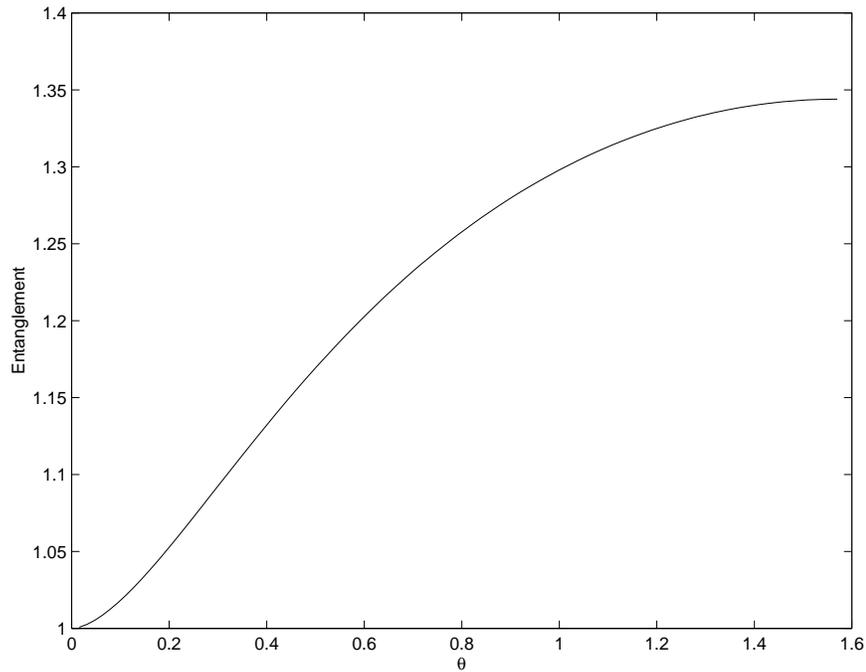}%
\caption{Entanglement between Alice and receivers in telecloning}%
\label{fig:tclone}%
\end{center}
\end{figure}
The entanglement between the two sides now increases with the overlap of the
two states $\left|  \psi_{1}\right\rangle $ and $\left|  \psi_{2}\right\rangle
$, but is always less than $\log(3)\approx1.585$, see Fig. (\ref{fig:tclone}).
However each qubit is maximally mixed so the entanglement between any one
qubit and the other three is one. This means that Alice's port qubit does
share a unit of entanglement with the other three qubits. This is consistent
with the requirement that perfect teleportation of two states employ a full
unit of entanglement. In this telecloning scheme, the amount of overall
entanglement is lower than in the universal case. It is an interesting
question whether a telecloning scheme giving the optimal two-state cloning
fidelity would also require less entanglement.

\section{Conclusion}

In this paper, we have shown the surprising result that a priori knowledge
makes no difference to the amount of entanglement required for perfect
teleportation. We have computed lower bounds for two-state teleportation
fidelity using a non-maximally entangled pure state as a channel, and the
exact result for the two-state fidelity with no entanglement.

This work opens a number of possible directions for future research. In this
paper, only pure entangled states were considered as channels for
teleportation. The investigation could be extended to mixed entangled states
also. The relationship between cloning and teleportation with a priori
knowledge could be investigated further by finding the amount of entanglement
required by a state-dependent telecloning scheme which preserves the
optimality of the clones produced. Asymmetric telecloning or general $N$ to
$M$ state-dependent telecloning could also be considered. It may be possible
to quantify exactly the relationship between the amount of information Alice
gains from her measurement, the amount of entanglement in the channel and
Bob's information. Our work provides a different way of understanding the
respective roles of classical information and quantum entanglement in the new
field of quantum information processing.

\textbf{Acknowledgments}

We thank Dagmar Bruss, Artur Ekert, Chiara Macchiavello, Mio Murao and Martin
Plenio for helpful discussions. L. Henderson acknowledges financial support of
the Rhodes Trust. L. Hardy thanks the Royal Society for funding.

\bibliographystyle{prsty}
\bibliography{two}
\end{document}